\documentclass{article}
\usepackage{graphicx} 
\usepackage{tikz}
\usepackage{amsmath}
\usepackage{array} 
\usepackage{authblk}

\usepackage{natbib}
\usepackage[english]{babel}  

\renewcommand{\cite}[1]{%
  (\citealp{#1} [\citenum{#1}])%
}

\title{The Negative Drift of a Limit Order Fill}
\author[1]{Timothy DeLise}

\affil[1]{Department of Mathematics and Statistics, Université de Montréal}
\date{January 2024}

\begin{document}

\maketitle

\begin{abstract}

Market making refers to a form of trading in financial markets characterized by passive orders which add liquidity to limit order books. Market makers are important for the proper functioning of financial markets worldwide. Given the importance, financial mathematics has endeavored to derive optimal strategies for placing limit orders in this context. This paper identifies a key discrepancy between popular model assumptions and the realities of real markets, specifically regarding the dynamics around limit order fills. Traditionally, market making models rely on an assumption of low-cost random fills, when in reality we observe a high-cost non-random fill behavior. Namely, limit order fills are caused by and coincide with adverse price movements, which create a drag on the market maker's profit and loss. We refer to this phenomenon as \textit{\textbf{the negative drift}} associated with limit order fills. We describe a discrete market model and prove theoretically that the negative drift exists. We also provide a detailed empirical simulation using one of the most traded financial instruments in the world, the 10 Year US Treasury Bond futures, which also confirms its existence. To our knowledge, this is the first paper to describe and prove this phenomenon in such detail.
    
\end{abstract}

\section{Introduction}
Financial markets employ limit order books to match buyers and sellers, producing transaction data that is then published to the world \cite{gould2013limit}. There are two basic types of orders: the market order and the limit order. Limit orders specify a price, a quantity, and a direction (buy/sell). Market participants send their orders to an exchange, and the exchange matches buy and sell limit orders against each other on a first-come, first-serve basis. Newly arriving orders are matched against the best available price of the previously existing orders. Counter parties to limit orders my give a price at least as good as the price of the limit order. Example: a limit order arrives to buy 1 lot at \$ 10. It is placed to wait for a matching order to arrive. Moments later another order arrives to sell 1 lot at \$ 10. These orders are matched, a transaction takes place, these orders have combined to produce a transaction. As orders arrive which don't have an immediate match, they get placed in what is called the \textit{\textbf{limit order book}} (LOB). The LOB is a list of all the outstanding, unfilled orders. It is organized according to the order attributes: \textbf{price}, \textbf{quantity} and \textbf{direction} (buy/sell). Orders which do not have immediate matches are said to \textit{add liquidity} to the market, since they will increase the amount of orders available. Market orders, and limit orders which have immediate matches are said to \textit{remove liquidity}.

Some dynamics of limit order books arise naturally. There is always a sell order with the lowest price, called the \textit{\textbf{best ask}}. Similarly the highest buy order is called the \textit{\textbf{best bid}}. As such, there are no resting buy orders with prices equal to or higher than the best ask, and vice versa for sell orders. The best bid and ask are where all of transactions occur, since they are optimal in their respective buy/sell categories. This rule arises naturally in limit order book markets. Incoming market orders and limit orders which remove liquidity are matched against the best available orders in the LOB.

\begin{figure}
    \centering
    \includegraphics[width=0.7\textwidth]{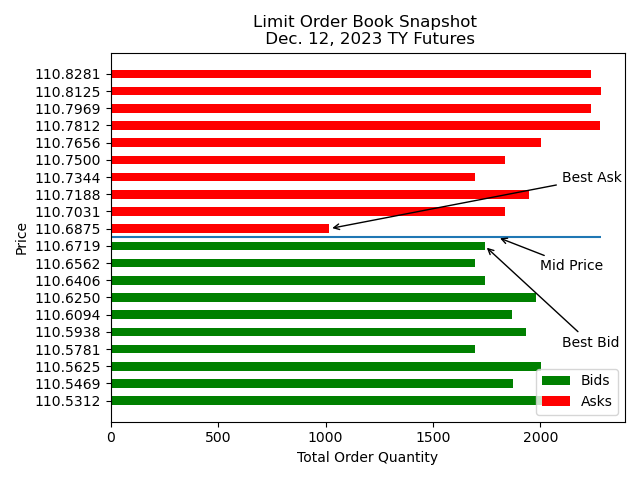}
    \caption[Limit Order Book Example]{In this figure we show a bar chart which represents the LOB data for an arbitrary time on Dec 12, 2023 from our Ten Year Treasury Bond futures (TY) data set. The data includes 10 levels (or distinct prices) above and below the midprice. All the price levels above the mid price correspond to sell orders, and vice versa for buy orders. At each price level we get a quantity, which is the total number of outstanding orders currently available at that price. }
    \label{fig:lob_schematic}
\end{figure}

Order prices cannot be specified to arbitrary precision. The price must be specified at discrete intervals according to what is called the \textit{tick size}. Many times the tick size of stocks is equal to 0.01, or one cent, but can be different, like 1/64 for prices in the 10-Year US Treasury Note futures. The prices produced by this discretization are called the \textit{\textbf{price grid}}. The basic information we discern from a snapshot of the limit order book is the quantity of orders available at any discrete price, see figure \ref{fig:lob_schematic} which shows a snapshot of the limit order book from the 10-Year Treasury futures market. 

Market orders specify a quantity and direction, but not a price. The market order will execute immediately at the best possible price. For a buy order, this price is the best ask price. It can be thought of as a limit order with a blank price field, the exchange fills it in once they check the limit order book for the best available price. There is a known cost involved in executing a market order, which is the cost of paying the market price. It is easy to visualize this cost with a \textit{round-trip trade}. A round-trip trade is a buy followed by a sell, or vice versa. If we buy something at the best ask and immediately sell it at the best bid, we will have lost an amount equal to 1 tick for each lot we transacted. We bought something and sold it immediately, and we lost money even though the market prices had not changed. In this light, it is easy to confirm that the slippage cost involved with a market order is equal to 1/2 the distance between the best bid and ask, which is usually equal to 1/2 tick. This cost is measured from the mid price. A buy market order is filled immediately on the best ask, which is 1/2 tick above the mid price. This is what we observe consistently in real markets.

The transaction costs associated with limit order fills are more difficult to analyze, since these orders are not filled immediately but are put in the order queue to wait until a matching order arrives. A motivating assumption is that, since market orders cost something to make, then the counter party to the transaction (the person who placed the corresponding limit order) should be paid the transaction cost, the 1/2 tick, by the market order. It makes sense too. A limit order placed on the best bid will get filled, assuming no other changes took place, at an equivalent to 1/2 tick compared to the mid price. However, the risk is that we cannot decide when the limit order is filled, and we must wait until it is our turn at the front of the queue. Subsequently, our order will be the best price available, it matches with a market order to create a transaction. So the real question here is: \textit{when will the order execute}?

Modern mathematics for high-frequency finance has endeavored to model the dynamics of limit order fills placed on the best bid or ask with probability distributions over time, such as the exponential distribution \cite{cartea2015algorithmic,contreras,avellaneda}. The underlying price process is usually a continuous stochastic process. The rationale here is that, the longer we leave our order exposed, the higher the chance of getting filled. Like this, limit orders still earn a 1/2 tick from transaction costs, the risk being that they don't get filled. While this is a nice beginning, it ignores a subtle truth in high-frequency markets: the price process is not continuous. The price process moves in discrete jumps from one price level to another. The minimum price change is 1 tick. If we place a buy limit order on the best bid, and the price moves down at all, it must move 1 tick. During that movement, the previous best bid becomes the new best ask. Now sell orders are present in the market which match the limit buy order, and an execution has happened.

This current paper is going to discuss these issues. There is intricate market behavior associated with limit order fills which creates a type of transaction cost. Indeed, this behavior has been touched upon in the market making literature under terms like \textit{adverse selection} \cite{cartea2015algorithmic,cartea2018}, where the mid price process is influenced by market order activity. The concept of adverse selection is related to the negative drift of this article. However, market making models in the literature still exclude that limit order fills are induced every time there is an adverse price movement, which is a key concept which should be appreciated.  

\subsection{The Negative Drift}

In high frequency financial mathematics we are often concerned with simulating order book fills in order to analyze strategies for market making or trading. Usually we are trying to use sophisticated mathematical or machine learning models to produce signals in real-time, creating a decision whether or not to place limit orders in the market, often at the best ask or best bid levels. In an example that we will use more in-depth later, the authors of \cite{cartea2015algorithmic} describe a stochastic optimal control problem, modeling the price process with stochastic differential equations and solving for the optimal control for two decision variables: whether or not to post limit orders on the best bid and/or best ask. Market orders are assumed to arrive randomly, and limit orders are assumed to be filled at a constant rate. Another popular paper which derives market making decisions using limit orders \cite{avellaneda} follows an assumption that the price process moves according to a continuous diffusion process, and limit orders are filled according to Poisson rates, without mention of any connection in the dynamics of market orders and the price process. Another popular research paper which this time uses machine learning and in particular deep learning to derive high frequency trade signals \cite{Zhang_2019} provides a trading simulation. It is assumed that we could operate a directional trading strategy by entering trades using limit orders and exiting trades with market orders, the penalty being that you may not get filled in time to enter each trade. Considering limit orders as a free replacement for a market order, where we can save the 1/2 tick slippage that we usually lose with a market order.

The common thread of the above research literature is the assumption that the movements of the price process are independent of the occurrence of limit order fills. It is as if the price is moving on its own accord, and limit orders are filled randomly at the best bid and ask without having any consideration of the price movement itself. But, what drives price movement? Could not order transactions have something to do with price process? Modern liquid markets have a bid-ask spread equal to 1 tick most of the time. As we will see more clearly later on, if the mid price moves at all, it must move 1 tick. When the mid price moves down the best bid instantly becomes the best ask, and vice versa for an upward movement. If there is a limit order on the best bid, and the price moves down at all, the best bid becomes the best ask, and any limit order will be filled with the matching orders that have arrived at the price level. This simple fact is missing from the current literature.

We will show that the there are particular consistent dynamics demonstrating that limit order fills are intimately connected and correlated with mid price movements. We analyze the chances of getting filled with a limit order posted at the best bid or best ask, according to actual markets. This kind of trading is also referred to as \textit{\textbf{at-the-touch}} market making in \cite{cartea2015algorithmic}. We contrast this with how limit order fills are usually assumed to work in mathematical models. We will show that common assumptions far too optimistic. They lead to unrealistic simulations and can even be misleading to a fledgling student in this topic of study. Instead, we suggest that there exist certain non-random dynamics in the market which should be staples of any high-frequency market model: 1) prices move in discrete full tick intervals, 2) there is \textit{\textbf{a consistent future negative drift in the mid price movement with respect to every limit order fill}}. This second point equates to a type of penalty associated with getting filled on a limit order that is very significant and consistent. The term \textit{\textbf{negative}} is in reference to the direction of the limit order itself. Buy order fills are accompanied by downward mid price movements and sell orders are accompanied by upward movements, on average. We propose that any high frequency finance order fill simulator needs to account for these drift dynamics. 

We support our claim with two main arguments, the first uses a very simple discrete market model with basic assumptions to prove theoretically that being filled on a limit order is on average accompanied by a move in mid price opposite the direction of the limit order. Secondly we perform empirical analysis by simulating a simple market making strategy in real time in the futures market and collecting statistics about each order and the market dynamics surrounding them. We utilize the trading platform, Trading Technologies, Inc. (TT), which has a sophisticated trading simulator that is very close to reality. With this combination of theory and practice, we provide strong evidence that our claims are true. We estimate the magnitude of negative drift encountered by limit order fills and find it to be substantial and consistent. The magnitude of the drift matches between the live simulation and our theoretical market model, as shown in table \ref{tab:expected-drift}.

The rest of the paper is organized as follows. In section \ref{section:existing-models} we will review some of the most popular limit order fill assumptions that have inundated the market making literate in recent years, illustrating the problem. Following this, section \ref{section:discrete-model} describes the discrete market model and the theoretical arguments for the negative drift of a limit order fill. Continuing from there we have section \ref{section:market-data-simulator}, which describes the live TT market making simulation and the results associated with it. Finally, we suggest a better way to simulate limit order fills for back testing purposes in section \ref{section:better-simulation}, with supporting empirical analysis.

\section{Existing Market Making Model Fill Assumptions}
\label{section:existing-models}

The first case we will discuss comes from a popular text book on high-frequency finance \cite{cartea2015algorithmic} that is used in some graduate courses. There are also numerous online resources associated with it, which provide working python notebooks for getting started.\footnote{We really commend the authors of this text book. Making working code available to students is really an amazing contribution, which helped to allow us to make the discovery of this current research. We also understand not all of the mathematics was written by the authors.} Specifically we refer to the stochastic optimal control problem for market making, chapter 10. There are a lot of good ideas here, which conclusions are drawn from sophisticated mathematical reasoning that makes sense. Unfortunately several assumptions lead the authors to analyze a market that doesn't exist. Their market does not resemble what we see empirically. Figure 10.3 in the text shows an example price path with best bid and best ask as a continuous process. Limit order fills occur seemingly at random on each side of the mid price. Using these assumptions we can create back test simulations using synthetic data that appear to be profitable, but in reality fail to produce a profit. We've identified that one of the main reasons for the discrepancy is the negative drift associated with order fills. 

For one of the market simulations, the authors assume that a limit order is filled against a market order with a probability of 1, in section 10.2.2, \textit{at-the-touch} market making. We will show later  that we find this probability to be much lower than 1 in practice. The more general idea presented in the chapter is to assume an exponential fill rate that depends on an exponential distribution and coincides with market order activity. This is generally better than assuming a 100\% fill rate, but is still inadequate, as we'll show later in our market making simulations. While the model is sophisticated, these assumption about high-frequency markets lead to a simulation which is overly optimistic.

More realistic assumptions are presented in section 10.4 of \cite{cartea2018}, where the incoming market orders push the mid price in a direction proportional to the market order activity. Presumably this should create a drag on limit order fills, since the fills should only be matched against market orders. However, the assumption here is that price movements themselves never reach the orders, and the market maker always is quick enough to move the order out of the way in time \cite{cartea2018}. Then, for at-the-touch market making, a 100\% fill rate is assumed, without price movements ever causing fills. The following points highlight the problems with these assumptions.

\begin{enumerate}
    \item The price moves according to a continuous processes that a real market never could in a ten second time span. The synthetic price process doesn't move in discrete jumps like it does in reality.
    \item There are market orders that presumably execute limit orders deep into the opposite side of the LOB, these orders can also move the mid price, but price movement never causes order fills. Thus the price process is still independent of limit order fills in unrealistic ways.
    \item Limit orders are never filled by adverse price moves, only by market orders. We quote \cite{cartea2018}: "\textit{Note that the agent continuously adjusts their posting relative to the mid price; hence, it is not possible for the mid price to move through the agent’s posts.}" We show in the current paper that this assumption is unrealistic, and that limit orders are indeed filled by adverse price moves 100\% of the time. 
\end{enumerate}

To summarize these issues, limit orders are filled randomly on each side of the book, the price is unaffected by these orders. Limit orders are never filled by adverse price movements, and the market maker comfortably nets a nice profit. We have found in experiments that when we use real market data and make more realistic assumptions, the simulated profit of such strategies turns negative. 

The second case has to do with the trading simulation of a deep learning model for predicting short-term alpha \cite{Zhang_2019}. In this study, the author's develop a deep learning model which can predict the future direction of the mid price of the limit order book: up, middle or down, at a high frequency and short time duration. For the purpose of a trading simulation, the authors convert these predictions to positions in real time, \textbf{executing their trades at the mid price with no transaction costs}. While the authors elaborate on why they know this assumption cannot describe a complete trading strategy, the logic is that they assume they can enter the trades with a limit order and exit with a market order, where the net transaction cost is zero. Hereby assuming that the result of their limit order was a positive 1/2 tick in transaction costs, canceling out the -1/2 tick imposed on the market order. What we show in our current work is that this assumption of the limit order fill in flawed in that every limit order fill incurs significant negative mid price drift on average. Moreover, the fill rate when the prediction is correct is much lower than the fill rate then when prediction is wrong. Indeed we will show that the order will get filled every time the mid price moves in an adverse direction, but only some of the time when the move is in a favorable direction, in ordinary markets.

The third and final case we will touch on here is also from a well-known high-frequency finance paper \cite{avellaneda}. Similar assumptions are made as in \cite{cartea2015algorithmic}, which we described above. Limit orders are placed above and below the mid price, and limit orders are filled by market orders at a Poisson fill rate. An effort is made here to account for market impact, where large market orders fill limit orders at multiple levels because they are so large. However, all of this order activity has no affect whatsoever on the evolution of the mid price, which is a continuous diffusion process. Again, these assumptions fall short of what we see in reality. Markets move in discrete jumps usually equal to one tick, where adverse movements fill orders more often than during beneficial movements. Again, while results are interesting and sophisticated based on the mathematics, they will be overly optimistic. 

What all three of these case studies reveal is that there is still some misunderstanding among the academic literature regarding the dynamics of limit order books and high-frequency markets. The following section introduces our discrete market model. It is intended to a be a minimal realistic model where we can showcase the dynamics of the negative drift using simple probability theory. Following that, we will explore our real-time simulation data, which supports our claims about the market model and the negative drift using real-time limit order fill simulations.

\section{Discrete Market Model}
\label{section:discrete-model}

In this section we will describe a simple discrete market model. It will be shown that without loss of generality, there exists a negative drift of the mid price with respect to limit buy order fills. A similar argument can be made for the case of sell orders. The following market model is not meant to be a perfect model. The purpose is to create a minimum sufficient, realistic model which can prove the existence of the negative drift. We feel the following model fits the mechanics of the market well enough where we can show our desired result under random conditions. The most important simplification made in this section is that the probability of upward and downward movements are the same. This allows us to illustrate the strict conclusion of the negative drift. These probabilities are usually very close in practice, which will be apparent in the following section with statistical analysis.

Our discrete model functions at discrete prices and time intervals. Real markets trade at discrete prices and order book events and transactions are instantaneous events which are more appropriately modeled with point processes than continuous processes. For the purpose of our simulation we will assume that the order book events arrive according to discrete time intervals. Thus, we start at time $t=0$ and $t \in \{0,1,2,...\}$. 

\begin{figure}
    \centering
    \includegraphics[width=\textwidth]{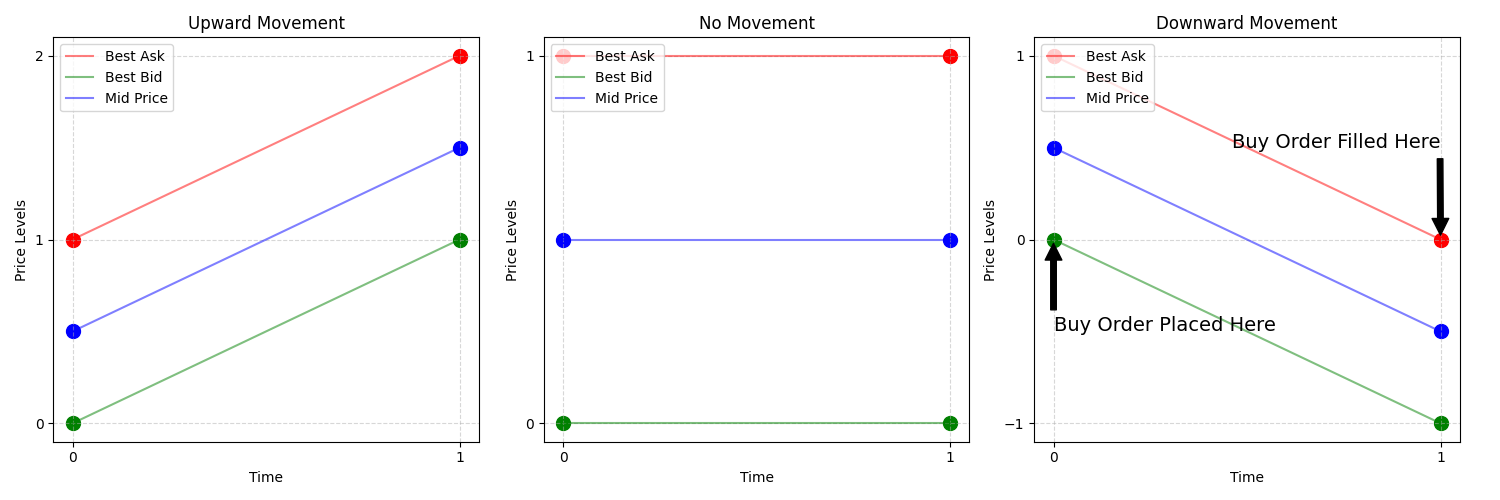}
    \includegraphics[width=\textwidth]{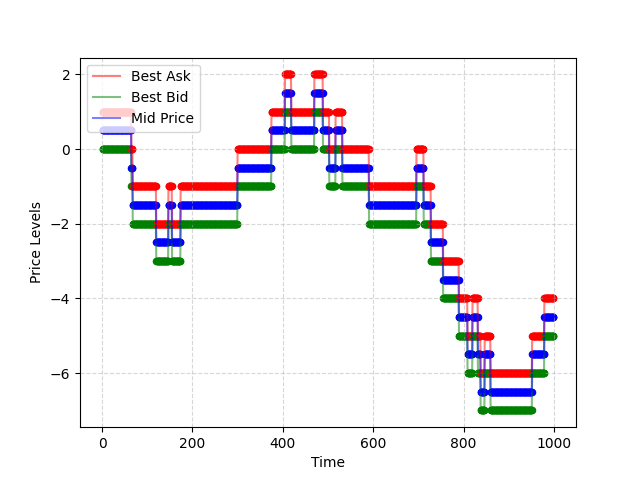}
    \caption[Discrete Market Model]{Above we show the 3 discrete events that can occur in our market model. Below we use probabilities $P(D) = P(U) = 0.017$ and $P(M) = 0.976$ and simulate 1,000 steps of this process. These probabilities were calibrated from the true 10 year bond futures market data, at 1 second time intervals.}
    \label{fig:up-middle-down}
\end{figure}

In this market model, we start where $t=0$ at an initial price, where the best bid and ask are separated by one tick. Each discrete time step gives rise to one of three possible random events: the prices move up one tick, the prices stay at the same value (middle), or the prices move down one tick (see figure \ref{fig:up-middle-down}). 

\begin{itemize}
    \item $U$ is the event of a + 1 tick movement.
    \item $M$ is the event of a 0 tick movement.
    \item $D$ is the event of a - 1 tick movement.
\end{itemize}

The sample space of our probabilistic experiment is $\{ U, M, D \}$. These three events are disjoint, the probabilities add to one, are all positive, and so define a discrete probability density. 

\begin{equation}
    P(U) + P(M) + P(D) = 1
\end{equation}

\begin{figure}
    \centering
    \includegraphics[width=\textwidth]{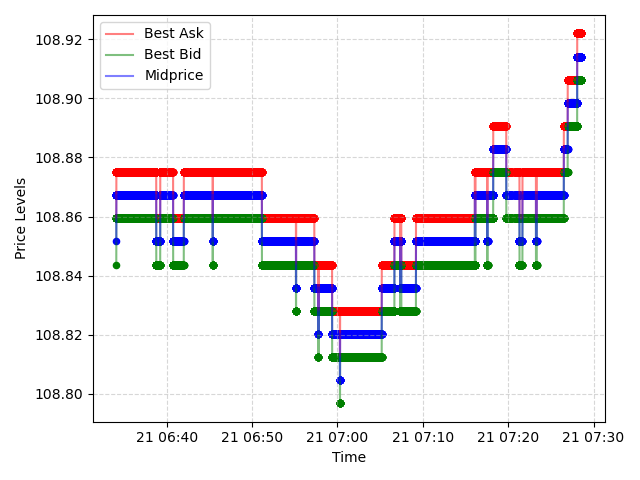}
    \caption[Example Ten Year US Treasure Bond Futures Data]{This is a small excerpt of the first 5,000 events we recorded in the TY data on November 21, 2023. While we have up to 10 levels of order book data, we plot the best ask and best bid with the mid price. One striking aspect is how similar this is to our discrete market model data. }
    \label{fig:zn-data-example}
\end{figure}

We simulate 1,000 steps of our simple market model and share this result in figure \ref{fig:up-middle-down}. We can compare visually to real market data, in figure \ref{fig:zn-data-example}, and see that it is hard to make a distinction between the two. Indeed, real markets move in discrete jumps and not in smooth continuous ways, which is especially important at very short time frames.

We wish to model the mid price process with respect to a limit order fill. To do this we are going to use conditional probabilities. The event we are going to condition on is the limit order fill, $f$. So, how is our order filled? At event time $t=0$, we enter the buy order to the limit order book on the best bid. During the course of each event, the order can be filled with Bernoulli probability rate $R_f$ regardless of any other market activity. We additionally impose the property that $R_f<1$. This means that the fill rate is not 100\%. We also must define an event where the price of the limit order equals the future best ask price, which occurs during a downward price movement and triggers the execution of a buy limit order. \textbf{\textit{It is precisely this event that is often omitted from market making literature, but which is one of the standard tenants of limit order book operation.}} Any buy order will be filled by a sell order at the same price for a matching quantity. When a downward price movement occurs, the best bid becomes the best ask, and any limit orders placed on the best bid must be filled before that price level becomes the best ask. See the top-right pane of figure \ref{fig:up-middle-down} for an example of how this happens. 

\begin{itemize}
    \item $f$ is the fill event.
    \item $P(f|M) = P(f|U) = R_f$ is the random Bernoulli fill rate for $M$ and $U$ events.
    \item $P(f|D) = 1$ the fill rate for a downward movement is 100\%.
\end{itemize}

We will now show mathematically, using expected values and conditional probabilities, that the average mid price movement, (\textit{the drift}), associated with a limit buy order fill is negative. This is also to say that the drift is negative with respect to the direction of the limit order, immediately after the order is filled. Let's define $d_t$ as the net change in mid price at time $t$.

\begin{itemize}
    \item $d_t$ is the change in mid price at time $t$.
\end{itemize}

And for a preliminary we show the average mid price movement as the expected mid price movement, using the law of total expectation. 

\begin{equation}
\label{eq:dm}
\begin{aligned}
    \mathbf{E}_t[d_t] &= (1) P(U) + (0) P(M) + (-1) P(D) \\
    \mathbf{E}_t[d_t] &= P(U) - P(D)
\end{aligned}
\end{equation}

Now we want to compute the expected mid price movement in the case where we can assume our limit order was filled, or $\mathbf{E}_t[d_t|f]$. Equation \ref{eq:dm} shows the drift over all the possible events that can occur in the market. The number of events in which our order would be filled is a subset of all the possible events, and is based on the fill rate. We assume the fill rates $P(f|M)$ and $P(f|U)$ apply to the middle and upward movements, however, the downward movements will always fill the limit order. The following is derived using the conditional probability formula, $P(A|B) = P(A \bigcap B)/P(B)$, for generic events $A$ and $B$ in some probability sample space.  

\begin{itemize}
    \item $P( U \bigcap f ) = P(f|U) P(U) = R_f P(U)$ the probability of a fill happening during an upward movement modeled as independent events. 
    \item $P( M \bigcap f ) = P(f|M) P(M) = R_f P(M)$ the probability of a fill happening during a middle movement modeled as independent events. 
    \item $P( D \bigcap f ) = P(f|D) P(D) = 1 P(D)$ the probability of a fill during a downward movement is $1$ by definition.
    \item $P(f) = P(D) + R_f( P(M) + P(U) )$ is the probability of getting filled. 
\end{itemize}

The last line uses the law of total probability, which states that if $\{ B_n : n = 1,2,3,... \}$ is a set of mutually exclusive and collectively exhaustive events (which we have), then the probability of an event $A$ is given by $P(A) = \sum_n P(A \bigcap B_n )$. The conditional expectation is the sum of the outcome of the events multiplied by the conditional probability of each event happening, 

\begin{equation}
    \label{eq:negative-drift}
    \begin{aligned}
    \mathbf{E}_t[d_t|f] &= (1) P(U|f) + (0) P(M|f) + (-1) P(D|f) \\
    \mathbf{E}_t[d_t|f] &= (1) \frac{ P(U \bigcap f) }{P(f)} + (0) \frac{ P(M \bigcap f) }{P(f)} + (-1) \frac{ P(D \bigcap f) }{P(f)} \\
    \mathbf{E}_t[d_t|f] &= (1) \frac{ R_f P(U) }{P(f)} + (0) \frac{ R_f P(M) }{P(f)} + (-1) \frac{ P(D) }{P(f)} \\
    \mathbf{E}_t[d_t|f] &= (1) \frac{ R_f P(U) - P(D) }{P(f)}
    \end{aligned}
\end{equation}

Since $R_f < 1$, and if we further assume that $P(U) = P(D)$, then we have shown that the expected mid price movement, given a buy limit order is filled, is always negative. Based on empirical data to follow, we find the real value of $R_f$ to be closer to zero than 1 in real markets. Also we find the empirical probabilities for $P(U)$ and $P(D)$ to be very close in value over one trading day in the 10 year bond futures, so these assumption are fine for our illustrative purposes here. The fact that $R_f$ is so small in practice makes the relative sizes of $P(U)$ and $P(D)$ somewhat insignificant. 

\subsection{Negative Drift in the General Compound Hawkes Process}

The previous discrete market model was chosen because it is simple but realistic enough to capture the essence of what is happening in real LOBs. It is a simple model but the important aspects function like a real market. Other existing discrete market models in the literature function similarly. For instance we can draw the same conclusion from the General Compound Hawkes Process (GCHP) model \cite{swishchuk2018general, sjogren2021general}, which models the mid price changes using a Markov chain. In this model there are only two possible states, up or down, which correspond to the direction of the previous mid price movement. The price always moves in discrete jumps equal to one tick, similar to ours. Instead of a middle state like in our model, a Hawkes process is used to model the inter arrival times of mid price changes, hence the name. The Markov probabilities can be referred to using the following notation. We further assume that all the probabilities are non-zero, forming a irreducible Markov chain transition probability matrix, meaning there exist steady state probabilities. We also define the drift in the mid price for this Hawkes process model as $d_t^h$, as follows:

\begin{itemize}
    \item $P(U|U)$ is the probability an up move follows an up move.
    \item $P(D|U)$ is the probability a down move follows a up move.
    \item $P(U|D)$ is the probability an up move follows a down move.
    \item $P(D|D)$ is the probability a down move follows a down move.
    \item $Q(U)$ is the steady-state probability that the process is in the up state.
    \item $Q(D)$ is the steady-state probability that the process is in the down state.
    \item $d_t^h$ is the change in mid price at time $t$.
\end{itemize}

We adopt the conventions of Markov chain mathematics and conditional probabilities in that $P(U|U) + P(D|U) = 1$ and $P(U|D) + P(D|D) = 1$. We compute the expected drift terms for the model, which depends on whether the current state is up $(U)$ or down $(D)$

\begin{equation}
\label{eq:hawkes-drift}
\begin{aligned}
    \mathbf{E}_t[d_t^h] = Q(U) \left( P(U|U) - P(D|U) \right) + Q(D) \left( P(U|D) - P(D|D) \right).
\end{aligned}
\end{equation}

Again, we have to assume that for a buy order, a downward movement will \textbf{always} result in a limit order fill. If the reader is not already convinced of this fact, please refer to the previous subsection. Using the law of total expectation we have first shown the expected drift of the price process itself in equation \ref{eq:hawkes-drift}. Remember that the outcome of an up movement is a +1 tick in mid price and that of a down movement is -1 tick. Like last time we use conditional expectation, conditioned on the fill event $f$, and a 100\% fill rate associated with the downward movement.

\begin{itemize}
    \item $P( X \bigcap f ) =  P(f|X) P(X) = R_f, \forall X \in \{ U|U, U|D \}$ the probability of a fill happening during an upward movement is $R_f$.
    \item $P( X \bigcap f ) =  P(X), \forall X \in \{ D|U, D|D \}$ the probability of a fill happening during a downward movement is 1.
    \item $P(f) = Q(U) \left( R_f P(U|U) + P(D|U) \right) + Q(D) \left( R_f P(U|D) + P(D|D) \right)$ is the probability of a fill event. 
\end{itemize}

 Finally, we derive the expected mid price movement given a limit order fill according the GCHP model, which conveys an analogous result to our simpler model above,

\begin{equation}
\label{eq:hawkes-limit-drift}
\begin{aligned}
    \mathbf{E}_t[d_t^h|f] = \frac{Q(U) \left( R_f P(U|U) - P(D|U) \right) + Q(D) \left( R_f P(U|D) - P(D|D) \right) }{ P(f) }.
\end{aligned}
\end{equation}

Equation \ref{eq:hawkes-limit-drift} can be re-written as

\begin{equation}
\label{eq:hawkes-limit-drift2}
\begin{aligned}
    \mathbf{E}_t[d_t^h|f] = \frac{  R_f (Q(U) P(U|U) + Q(D) P(U|D) ) - (Q(U)P(D|U) + Q(D) P(D|D) ) }{ P(f) }.
\end{aligned}
\end{equation}

If we make the simplifying assumption that the steady state probabilities of upward and downward movements are the same, that $Q(U) = Q(D)$ (which is realistic and sufficient for our purpose), then we have 
\begin{equation}
\begin{aligned}
    \mathbf{E}_t[d_t^h|f] = \frac{  R_f (P(U|U) + P(U|D) ) - (P(D|U) + P(D|D) ) }{ P(f) }.
\end{aligned}
\end{equation}
With the simplifying assumption that steady state probabilities are equal we can quickly conclude that $P(U|U) + P(U|D) = P(D|U) + P(D|D)$. Combining this with the assumption that $R_f < 1$ allows the conclusion that $\mathbf{E}_t[d_t^h|f]<0$. In practice, the precise probabilities governing upward and downward movements are not all that important for this, since upward and downward probabilities are roughly equal, but the magnitude of $R_f$ is very small compared to the magnitudes of upward and downward price movements.

The point here is that this negative drift should arise in any market model which recognizes the realistic assumption that a buy limit order placed on the best bid will be filled when the price moves at all in the opposite direction (down), but only sometimes when the price moves in the same direction (up).  During these adverse movements, a buy order will be executed \textbf{on the best ask}, and not the best bid as originally hoped for. The market has shifted against the trader's new position. Thus, this is equivalent to a form of slippage that we call \textbf{\textit{the negative drift}}.

\section{Live TT Simulator}
\label{section:market-data-simulator}

Some readers may be a little reluctant to trust the theoretical findings of the previous section. Indeed this notion strays from the usual assumption that market makers passively place their orders and randomly get filled, netting the difference between their buys and sells, which usually is about 1 tick. That assumption is not good enough. Real markets do not behave like that, and getting a limit order filled also incurs a cost, the negative drift. 

In this section we will provide further evidence through statistical analysis of a market making simulation in the 10 year US treasury bond futures market, ticker symbol TY. We have implemented a naive market making strategy in real time using our own data pipeline with the Trading Technologies, Inc (TT) platform and our own data infrastructure. Our data pipeline receives, processes and saves the real-time streaming tick data directly from the provided .NET API. The data stream consists of market depth (level 2) and time \& sales (level 1) data. Data values arrive asynchronously with inter arrival times between events often a small fraction of a second. The arriving data is simultaneously published to a message relay software, RabbitMQ, and saved to a PostgreSQL database. In this way we can use the data in real-time and also for historical analysis and backtesting. The TT software offers a simulation environment that allows the users to simulate limit orders that are matched against real market activity in real time. Market orders and queue position sizes are used, among other things, to simulate every limit order fill as accurately as possible. This is not a back test, but a real-time simulation where orders executions happen at the whim of the markets, as realistically as possible. 

We provide a case study which takes place between 6AM and 1PM EST on November 21, 2023 for the December, 2023 expiry. We implement a naive market making strategy which posts limit orders for 1 lot on the best bid and best ask to buy and sell respectively in the TY futures. If the position is different from zero, meaning that the trader is long or short 1 lot, the trader will only post the order that will return the system to a zero position. For example, if the position is long 1 lot (meaning the position is 1), the trader will only post the sell order at the best ask. Once the order is filled and the position returns to zero the trader will then post both buy and sell orders again. This is the entirety of the strategy. The robot trader passively gets filled on buy and sell limit orders, always posted on the best bid or ask. While this is a naive strategy, the placement of the orders on the best bid and ask is standard practice in the market making literature \cite{cartea2015algorithmic}. The dynamics gleaned from this analysis are meant to apply to general markets for average limit order fills.

During the course of the simulation we are recording every order book event, every transaction, and every action concerning our simulated orders. These events are fed back from the exchange. Concerning our simulated orders, there are three basic actions that can occur: \textit{order added}, \textit{order canceled}, and \textit{order filled}. With the details and timestamps of these actions and events we can follow along and compute statistics of the data surrounding our trades. Concerning the quality of the simulated order fills, the TT platform's simulation environment is sophisticated. It models the orders' positions in the order book queue and always matches the orders against other real orders in the market. While no simulation can match reality perfectly, we feel this live market simulation is very accurate. We have compared many simulated orders with live orders and found executed prices are very close in each case.

\begin{figure}
    \centering
    \includegraphics[width=\textwidth]{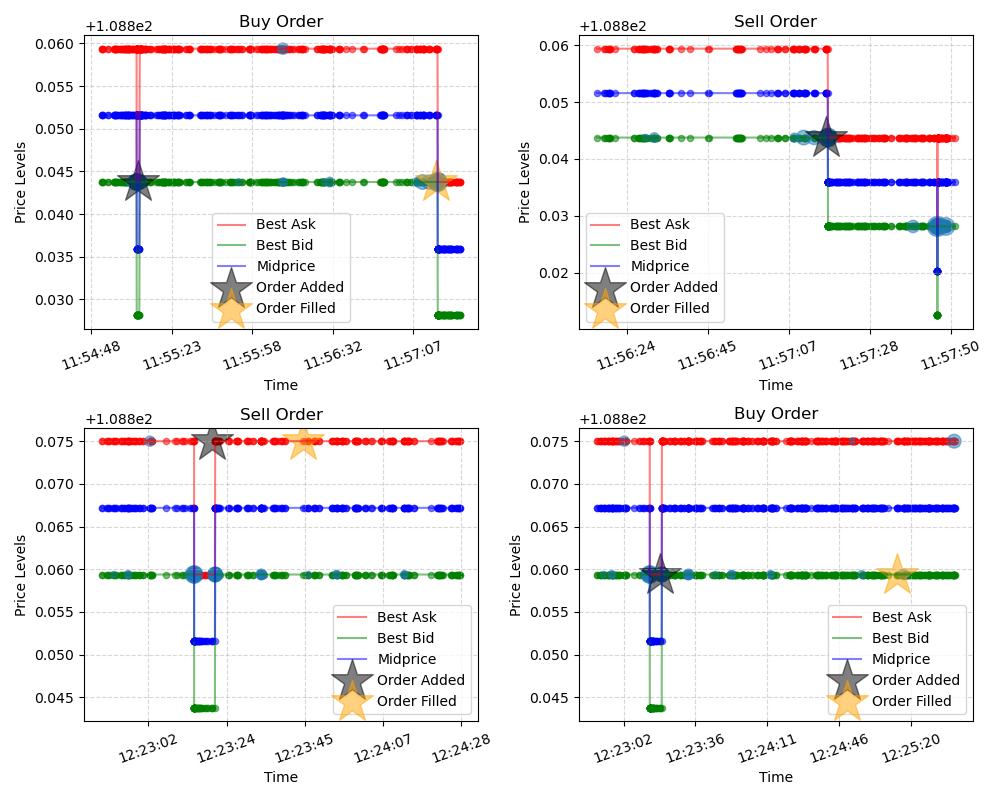}
    \caption[Visualize Limit Order Fills]{Here we visualize 4 individual orders over their lifetimes. In the first frame (top left), we see a buy order placed on the best bid, indicated by the black star. The order is filled at the location of the orange star. Notice that this order fill coincides with a drop in the mid price, where the best bid becomes the best ask. The subsequent trade (top right) is a sell order, and we see that this order is never filled as the price moves down. The bottom two orders show the case where the order is filled without the adverse price move. In both of these cases the order is filled while the price level remains unchanged. We also plot the market order transactions in blue, size representing the quantity of the order. Each data point is mid price movement during 100 events following each fill. }
    \label{fig:order-breakdown}
\end{figure}

Our simulation contains a total of 1,683 orders. Each order consists of 2 events. The first event is the order added event and the second event is either the order filled or the order canceled event. Not every order gets filled, and if the price moves away from the limit order, a new order at the new price level is entered as a replacement. In figure \ref{fig:order-breakdown} we break down four of the orders from the early part of the day. Interestingly these few orders showcase negative drift in action. We see that the first order fill coincides with one of these adverse price movements, where the best bid becomes the best ask instantly. The second order (top right) is a sell order that never gets filled before the price has moved away from the order. The third and forth (bottom row) example orders show the case that benefits the market maker: when the orders are filled without an adverse price movement. We also plot the market order activity in blue, since later on we use these market orders to help simulate random limit order fills.

The bottom-line statistic that we are looking to show is the negative drift associated with a limit order fill. Then we will continue to estimate the probabilities and fill rate that could be used to calibrate our market model and compute the average drift in equation \ref{eq:negative-drift}. Finally we will shed light on the more intimate dynamics at play between market order activity, order fills, and price movement. 

Since we are concerned with the price behavior following order fills, we target these events in our data, which are defined by a high-resolution time stamp. We then record the mid price movement over a window starting at this time stamp and extending for 100 events into the future. For every order fill event, we will record one mid price movement, and then we will group the data into two groups, depending on whether the order was a buy or a sell. We then randomly sample a similar number of data points as a control. 

\begin{figure}
    \centering
    \includegraphics[width=\textwidth]{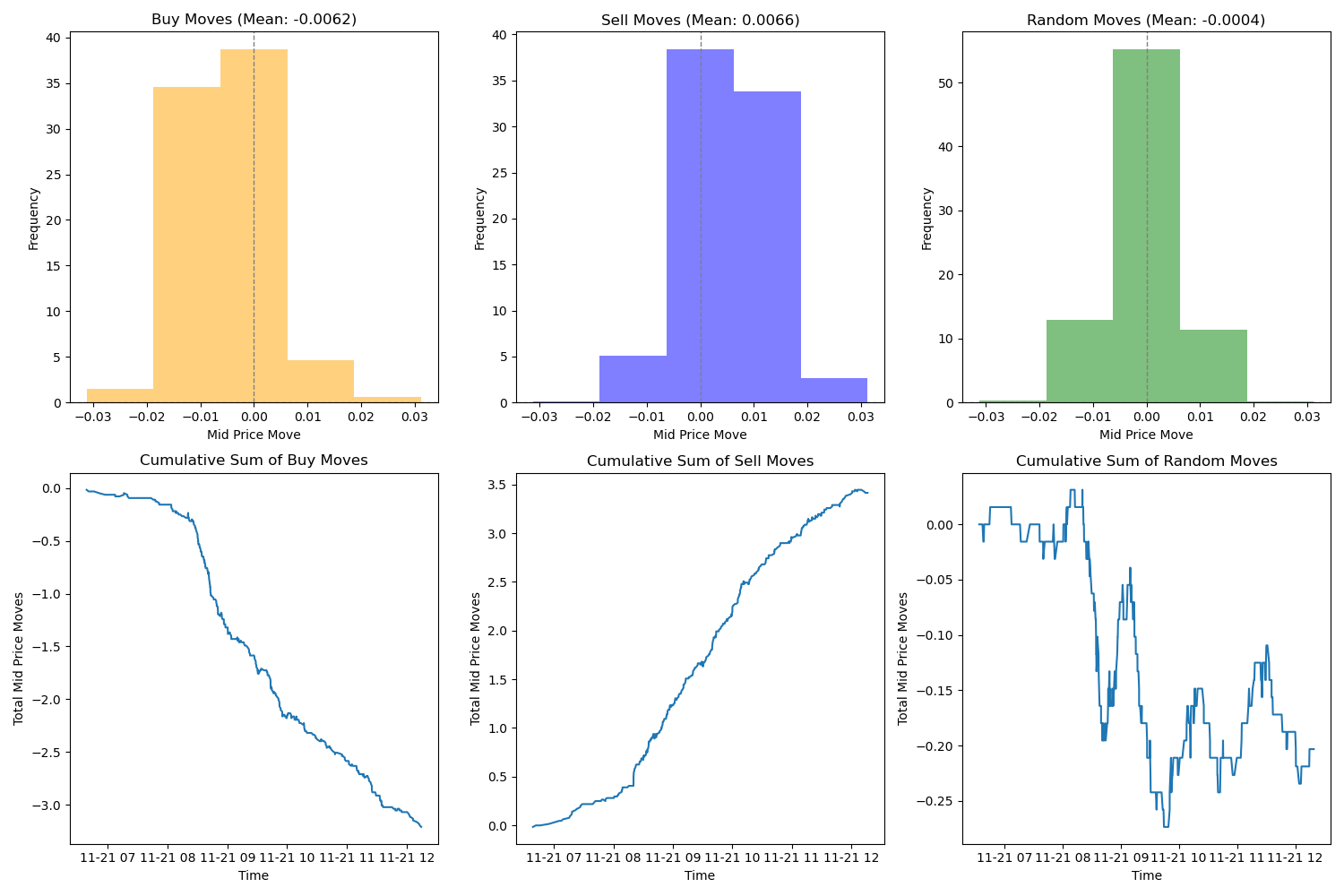}
    \caption[Price Drift Following Limit Order Fills]{In this figure we plot the results of our analysis on the mean drift of the mid price following a limit order fill. The top row shows histograms and the bottom row shows the cumulative sum of the same data. The first column is the data for mid price moves following buy orders, the second column corresponds to sell orders, and the third column shows mid price moves following randomly-selected time points.}
    \label{fig:drift-combined-plots}
\end{figure}

The results of our analysis are shown in figure \ref{fig:drift-combined-plots}. Each sample consists of 521 data points. The trend is clear. There is a significant downward price drift following buy order fills and an equally significant upward drift following sell order fills. Compare this to the control sample which shows hardly any net directional drift at all. In each case the direction of the drift is opposite the direction of the order. On top of this we have computed the average drift in each case, which comes out to approximately $-0.0065$. To put this metric into perspective, the tick size in the TY futures is $1/64 = 0.015625$. The traditionally accepted rule is that a \textbf{market order} always incurs a slippage cost penalty equal to 1/2 of a tick compared to the mid price, or $0.0078125$. This is because with a market order we are allowed to fill our order immediately against the best bid or best ask on the opposite side of the book, which is usually 1/2 tick away from the mid price. It was also traditionally presumed that a limit order could be used to avoid this penalty by using passive orders on the same side of the book. The risk being that the order isn't filled immediately. What we have shown here is that a limit order fill does incur a type of slippage penalty, and the magnitude of that penalty is shown to here to be very close to 1/2 tick. Furthermore, according to our simulation, 1/3 of orders were not filled at all. This is to say that we incurred a negative drift of $-0.0065$ on all of the filled orders, but these orders only account for 2/3 of our orders. The remaining 1/3 of the orders were unfilled. The price moves away from them. In order to make a transaction we would need to re-adjust our limit price in a non-favorable direction. This case is illustrated in the top-right subplot of figure \ref{fig:order-breakdown}. This explains why the assumptions of the simulation \cite{Zhang_2019} are so troublesome: negative drift affects filled orders, but many orders are unfilled, especially during times when the prices move away in a favorable direction.

\subsection{Calibrating the Discrete Model}
\label{section:calibrating-discrete-model}

We now estimate the values of the transition probabilities and fill rate of our discrete market model from the empirical data: $P(U)$, $P(D)$, $P(M)$, and $R_f$. For this calculation we will normalize our time series so that it has uniform time intervals similar to the assumptions of the discrete model. For this we choose a time resolution of 1 second intervals. We keep track of the information pertinent to calculating the transition probabilities and fill rate. For each discrete time interval we record the most recent best bid, best ask and mid price, as well we record whether or not an order was active during the time period. We also record whether or not an order was filled during the time period.

In order to compute the probabilities, we take the re-sampled data, which now consists of 20,743 1-second time intervals. Resampling the data to larger uniform time intervals serves to simplify the analysis a bit. Indeed with the rapidly changing events of the pure tick data, it is a little harder to capture the essence of the negative drift. At one second uniform time intervals, it is clear what is happening. It is also standard practice to simulate limit order books at uniform time intervals, even if it is not perfectly realistic. For the re-sampling we take the most recent bid and ask prices, as well as whether or not orders were present and filled during each interval. Later on in the section on simulating orders, we will show how this analysis can also work for non-resampled data. 

We compute the mid price changes between time steps. We sum the number of times the price increased and divide by the total number of time steps to get $P(U)$. We sum the number of times the price decreased and divide by the total number of steps to get $P(D)$. We take the number of times there was no movement and get $P(M)$. These are are empirical probabilities. We present the findings in table \ref{tab:probabilities}.

\begin{table}
    \centering
    
    \renewcommand{\arraystretch}{1.5} 
    \begin{tabular}{|c|c|}
        \hline
        \textbf{Statistic} & \textbf{Value} \\
        \hline
        $P(U)$ & 0.0173 \\
        \hline
        $P(D)$ & 0.0173 \\
        \hline
        $P(M)$ & 0.965 \\
        \hline
        $R_f$ & 0.018 \\
        \hline
        $P(f|D)$ & 0.99 \\
        \hline
    \end{tabular}
    \caption{Statistics for the discrete model calibrated from the market data.}
    \label{tab:probabilities}
\end{table}

The fill rate of a limit order in the face of an adverse price movement is assumed to be 100\% in our discrete model. And $R_f$ is assumed to be the rate at which the orders are filled in the cases when the price movement was not adverse. Here we assume a Bernoulli distribution with probability $R_f$ of getting a fill during each time interval. In order to test these assumptions we have computed the probability of a fill during adverse price movements by counting the number of times we observe an adverse price movement while a limit order was placed opposing the movement. This includes when we have a buy order placed during a downward price movement or when we have a sell order placed during an upward movement. Then we check what portion of these events resulted in a limit order fill. This is an empirical calculation of $P(f|D)$ for buy orders according to the discrete model. The conclusion is that, in fact, these orders are filled 99\% of the time. Alternatively we compute empirically $R_f$ by considering what portion of events we had limit orders placed but the price did not move in an adverse direction. The portion of these events that resulted in limit order fills is thus $R_f$. The results are presented in table \ref{tab:probabilities}.

Finally, we can plug these probabilities into equation \ref{eq:negative-drift} to compute the expected drift of a limit order fill $\mathbf{E}_t[d_t|f]$ after tuning our model to the real data. We then compute ${\mathbf{\hat{E}}_t}[d_t|f]$, which is the empirical average drift of the mid price in the time interval following and order fill. These two results come out to be very close, indicating our model is calibrated well, and the phenomenon really exists. 

\begin{table}
    \centering
    \renewcommand{\arraystretch}{1.5} 
    \begin{tabular}{|c|c|}
        \hline
        \textbf{Statistic} & \textbf{Value} \\
        \hline
        $\mathbf{E}_t[d_t|f]$ & -0.48 ticks \\
        \hline
        ${\mathbf{\hat{E}}_t}[d_t|f]$ & -0.45 ticks \\
        \hline
    \end{tabular}
    \caption{Theoretical and empirical drift following order fills. Numbers presented in number of ticks.}
    \label{tab:expected-drift}
\end{table}

\section{A Better Way to Simulate Limit Order Fills}
\label{section:better-simulation}

The goal often arises in quantitative finance to \textit{back test} trading strategies. \textit{Back testing} is the practice of simulating a trading strategy using historical data in order to understand how such a strategy behaves given the history of the market. Back testing is paramount to quantitative finance. If done correctly, it can give insight to the behavior of trading strategies without the time cost of live simulation. If done incorrectly, it can mislead investors to invest in unrealistic endeavors. 

Existing literature posits certain assumptions for how to simulate fills in a market making context, as we've discussed previously. Usually these involve a certain fill rate matched against market orders, where limit orders are filled against incoming market orders according to an exponential distribution. Simulations which create and use synthetic data have their own difficulty of simulating the price movements and market orders. In our case we are backtesting, which means we use real historical market data. However we still would like to simulate limit order fills in order to understand how an algorithm would have performed over the history of the market.

In this section we develop simple rules for simulating limit order fills with historical data, based on our discrete model above, and the negative drift. The result can be the backbone of a simple and accurate back tester for market making strategies. To serve as a benchmark, we also use the TT live simulator to simulate the naive market making strategy over the same time period as the back test. The goal is for our simulated back test to match the live TT simulator as well as possible. 

The experiment in this section consists of 3 back test simulation techniques, compared to the TT live simulation as baseline. The first two techniques are described in \cite{cartea2015algorithmic, cartea2018}, and suggested to be feasible in the market making literature. The first technique assumes that limit orders placed on the best bid ( or ask ) will be filled by sell market orders ( or buy market orders ) at a 100\% fill rate. This is to say that our limit order will always get filled by the first arriving market order. The second technique is referenced in \cite{cartea2015algorithmic} and \cite{avellaneda}, which assumes fills arrive according to an exponential distribution, with the rate, $\lambda_f$, calibrated from the market data. According to this model, a limit order is filled against the first market order with an inter-arrival time greater than or equal to a realization from an exponential random distribution with calibrated $\lambda_f$ parameter.  The third technique is our new technique which assumes that limit orders placed on the best bid (ask) will be filled by adverse downward (upward) price movements. Besides this, the orders which do not face adverse price movements are filled according to a Bernoulli distribution with the rate, $R_f$, calibrated from the market data, as described in section \ref{section:calibrating-discrete-model}.

\begin{enumerate}
    \item \textbf{Back Test Technique 1:} 100\% fills against market orders.
    \item \textbf{Back Test Technique 2:} Exponential fills against market orders.
    \item \textbf{Back Test Technique 3 (Ours):} Adverse fills and Bernoulli rate.
    \item \textbf{Baseline:} Live simulation.
\end{enumerate}

In order to take the first step in comparing the aforementioned backtesting techniques, we calibrate our models and then perform the simulation on the same data. This goes against a common practice in statistics and machine learning, which calibrates a model in-sample, and evaluates out-of-sample. The reason we chose to do this is to highlight the assumptions of techniques 1 \& 2. In-sample, the calibrated models do not fit the data, as well will see. So, why should we expect these models to perform well out-of-sample. 

The data comes from the the time between 8AM and 5PM EST, November 30, 2023 in the 10-year bond futures' March 2024 contract. For this experiment we do not down sample the data at all, but use the raw tick data, event by event. We do this to get the most accurate back test possible. The naive market making strategy that we employ is to post buy and sell orders on the best bid/ask according to the current position of the market maker. If the position is zero, both buy and sell orders are placed. If the position is +1, only sell orders are placed. If the position is -1, only buy orders are placed. Order quantities are always 1 lot. In this way the market maker's position oscillates between 1 and -1 for the duration of the test. This is the same strategy set up as before.

\begin{table}
    \centering
    
    \renewcommand{\arraystretch}{1.5} 
    \begin{tabular}{|c|c|}
        \hline
        \textbf{Item} & \textbf{Value} \\
        \hline
        Contract & TY Mar24  \\
        \hline
        Time Period & 8AM - 5PM EST, 11/30/2023  \\
        \hline
        Number of Events & 576,670  \\
        \hline
        P(U) & 0.00186 \\
        \hline
        P(D) & 0.00187 \\
        \hline
        P(M) & 0.9962 \\
        \hline
    \end{tabular}
    \caption{Live Simulation Data Description}
    \label{tab:data-desc}
\end{table}

For technique 1, no model calibration is required, since the fill rate is simply 100\%, matching limit orders against incoming market orders. For technique 2 we calibrate the exponential rate by using the inverse of the mean inter arrival time for order fills from the TT simulation, relying on the assumption of the exponential distribution. We then match limit orders to incoming real market orders if the inter arrival time meets or exceeds random exponential arrivals with rate $\lambda_f$. For technique 3 (ours) we calibrate the model and fill rate $r_f$ as described in the previous section, to the live simulation. Summary statistics from the data are provided in table \ref{tab:data-desc}, calibrated fill rates in table \ref{tab:fill-rates} and we display the number of orders and fills for each simulation in table \ref{tab:orders-and-fills}.

\begin{table}
    \centering
    
    \renewcommand{\arraystretch}{1.5} 
    \begin{tabular}{|c|c|c|}
        \hline
        \textbf{Technique} & \textbf{Parameter} & \textbf{Value} \\
        \hline
        2 & $\lambda_f$ & 0.0842 \\
        \hline
        3 & $R_f$ & 0.000625 \\
        \hline
    \end{tabular}
    \caption{Back Test Rates}
    \label{tab:fill-rates}
\end{table}

\begin{table}
    \centering
    
    \renewcommand{\arraystretch}{1.5} 
    \begin{tabular}{|c|c|c|c|}
        \hline
        \textbf{Technique} & \textbf{No. of Orders} & \textbf{No. of Fills} & \textbf{Global Fill Rate} \\
        \hline
        1 & 13,935 & 11,859 & 85\% \\
        \hline
        2 & 3,785 & 1,134 & 30\%  \\
        \hline
        3 & 3,459 & 2,259 & 65\%  \\
        \hline
        Baseline & 3,305 & 1,985 & 60\% \\
        \hline
        
    \end{tabular}
    \caption{Orders and Fills}
    \label{tab:orders-and-fills}
\end{table}

In figure \ref{fig:pnl-comparison} we present the mark-to-market P\&L for the four techniques over our experimental time period. Reiterating that all four of these trials are using the same market data to simulate order fills.

Technique 1 enjoys a 100\% fill rate with arriving market orders and orders are never filled during adverse price movements. Orders are filled often and are able to be replaced quickly. The effect of this is a high number of total orders and fills compared to the baseline. The high fill rate creates a great profit path, in blue.  This ranks as the most unrealistic case, since the P\&L diverges from the baseline the most.

\begin{figure}
    \centering
    \includegraphics[width=0.65\textwidth]{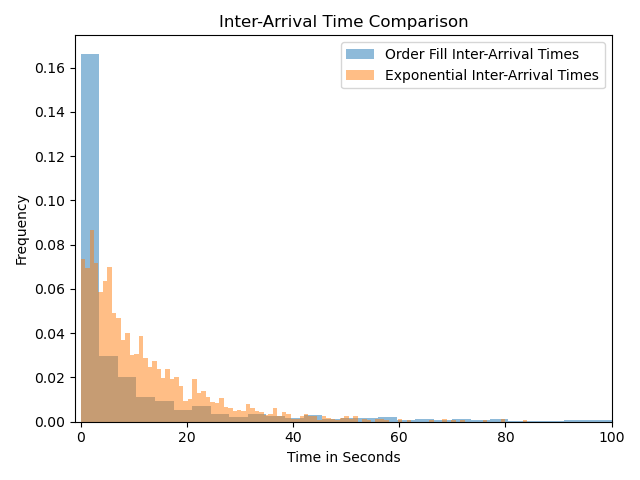}
    \caption[Heavy Tailed Inter-arrival Times of Order Fills]{In this figure we compare the distributions of the inter-arrival times of filled orders according to the live simulation with synthetic exponential inter-arrivals after calibrating the mean to match real data. There were a total of 1,985 filled orders during this simulation, with a mean inter-arrival time of 11.9 seconds. These are used to create the blue histogram. The exponential inter-arrival times here, in orange, have the same mean time, and we've simulated the same number of them. If the actual inter-arrival times followed an exponential distribution, these histograms should fit together closely. However we see that there are much more shorter and longer times and less medium-range times in reality than implied by the exponential distribution. We have reduced the visible range of the x-axis here because the largest inter arrival from live simulation was greater than 250, which made the histograms difficult to see. }
    \label{fig:iit}
\end{figure}

Technique 2 matches limit orders to market orders according to an exponential random variable. While the total number of orders is closer to the baseline, the number of fills is much lower. We attribute this to the misspecification of the distribution of inter-arrival times of orders. The inter-arrival times of order fills do not appear to fit an exponential distribution, see figure \ref{fig:iit}. This discrepancy has been pointed out in previous literature \cite{swishchuk2018general}. Instead of exponential, the shape of the distribution of inter-arrival times of orders may fit better to heavy-tailed, Pareto like, distributions. If we try to simulate fills using real data, but assuming exponential inter-arrival times, the mean no longer matches. In reality more inter-arrival times are short, but the few that are long are very much longer than suggested by the exponential distribution. The effect is that we have fewer fills, but the fills we get have a higher mean inter-arrival time, reducing the number of total fills. Nevertheless, in the absence of the adverse price movement fill logic, this strategy would still make good money even with a low fill rate. Comparing the P\&L with the baseline, real market fills aren't nearly so generous. 

Technique 3, which combines adverse fills and a calibrated Bernoulli fill rate attains the closest global fill rate and also P\&L to the baseline model. In particular, in the top-right pane of figure \ref{fig:pnl-comparison}, we see a section in the early part of the chart has a quick draw down in P\&L for the baseline model. Our simulation catches that trend perfectly, most likely due to the adverse fill logic. The remaining discrepancies in the P\&L chart between technique 3 and the baseline indicate that there is still room for improvement. The result of our back test technique is much closer to the baseline in terms of P\&L and also number of orders, fills, and fill rate than the other two techniques.

\begin{figure}
    \centering
    \includegraphics[width=\textwidth]{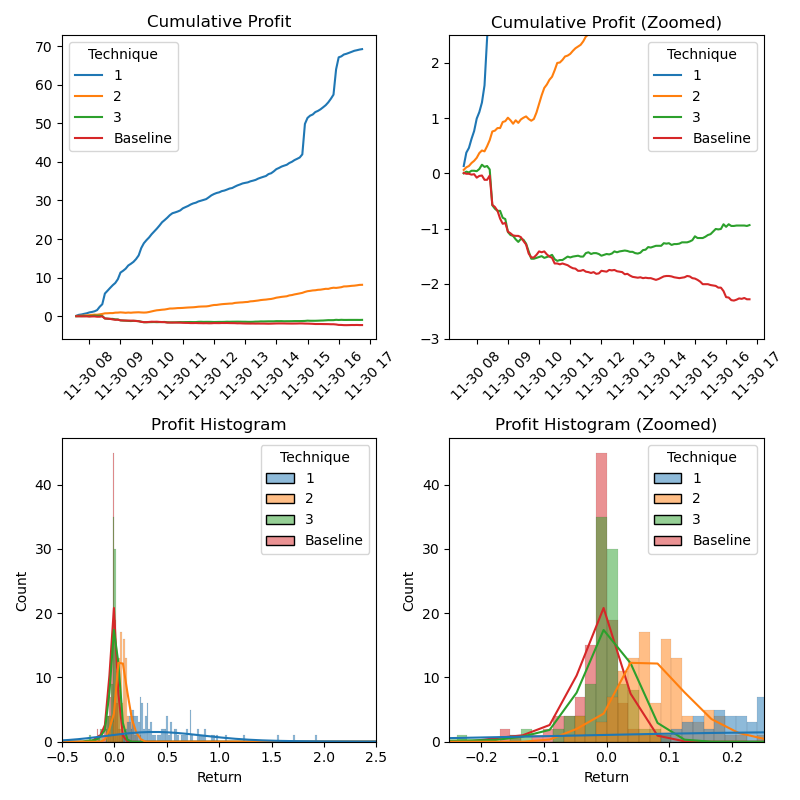}
    \caption[Back Test Result Summary and Comparison]{In this figure we summarize the results of the back test experiment. All of these images are based on the cumulative mark-to-market profit and loss over 5 minute windows throughout the day. On the top row of images, we plot the cumulative profit and loss over the day, with the right pane zoomed in to show the details around the baseline. The bottom row of images shows the data in histogram form, again, zoomed on the right hand side to show better the details surrounding the baseline performance. Indeed, the profit from the 100\% fill rate assumption creates a return stream that is on a completely different scale than the rest. Our technique (\#3) matches most closely with the baseline target by a wide margin. }
    \label{fig:pnl-comparison}
\end{figure}

\section{Discussion}

Even though topics like \textit{adverse selection} have been discussed in the market making literature \cite{cartea2015algorithmic}, no existing study has previously described the phenomenon of the negative drift associated with a limit order fill as succinctly as this current paper. We have described with simple theoretical models and replicable empirical results that the phenomenon exists. The conclusion is based on a simple assumption regarding a limit order placed on the best bid or ask, that it will be filled by adverse price movements 100\% of the time in basic circumstances. We have shown that this assumption is logically accurate and actually observed in financial markets.

An argument could be made that this conclusion is based on actions of the naive (uninformed) trading strategy, while an informed trader will be able to move or remove their limit orders in time to avoid adverse price movements, by utilizing what is called a \textit{short term alpha} signal. While it is beyond the scope off this current article, we can make some comments. In practice no short term alpha signal is completely reliable. We have experimented with order flow, machine learning, and Markov chain-based short-term alpha signals. We find that indeed such signals can predict short-term alpha (the next mid price movement direction) to a significant extent. However financial data is noisy, and even good predictions have a correlation with the future mid price movement of only 15 - 25 \%. This is to say that many of our predictions will be wrong, and we will be unable to avoid every adverse price movement. The negative drift still affects such strategies heavily. While there is more work too be done in this area, we believe that our negative drift model should be the default setup which we use to evaluate short-term alpha signals. Using a simulation strategy as described in the previous section, we should be able to evaluate methods that aim to avoid the negative drift. 

We showed in section \ref{section:discrete-model} that the negative drift can be modeled in the GCHP model \cite{swishchuk2018general} when we assume limit orders are filled by adverse price movements and a Bernoulli rate for non-adverse movements. We hypothesize that the remaining discrepancies between our back test implementation and the live simulation would be narrowed by assuming that inter-arrival times of non-adverse fills follow a Hawkes process \cite{hawkes} instead of a Bernoulli or exponential fill rate, which is an assumption gaining more traction in recent research \cite{kirchner, swishchuk2018general, sjogren2021general}. It may make sense to use the GCHP model as the basis for the simulation procedure. Indeed the GCHP model enjoys some nice mathematical properties that have been developed recently \cite{swishchuk2018general}, such as the law of large numbers and functional central limit theorem. However, these results are based on limit theorems that effectively convert the GCHP to a continuous diffusion process, which leads to incorrect specification of the high-frequency price process, leading us in the wrong direction.

We chose the present set up because it is simple yet expressive. We define simple rules and find that the outcomes match reality much more closely than any previous market-making model for limit orders. The richness and accuracy the simulations can provide based on such simple rules is inspiring. Simple building blocks can give rise to rich features and interesting patterns when repeated many times \cite{MAN83}. It reinforces our confidence in the findings when the simple model creates dynamics that are indiscernible from nature.

Markets are discrete systems, where prices jump from one price to another on the price grid. If a price moves at all, it moves at least one tick. Moving one tick is enough to cause a fill of a limit order which is placed on the best bid or the best ask in liquid markets. This should be a staple of any high-frequency market making model in limit order books. Surprisingly, these basic elements have been absent or under-represented from previous market making and high-frequency finance literature until now. We took the step to incorporate these assumptions into a mathematical market model. When we did that, we discovered and defined the negative drift of the mid price with respect to a limit order fill in theory and practice.

\bibliography{main}
\end{document}